# 2021 Snowmass Process

## *Fe-based high field superconductors for cost-effective future accelerator magnets*

F. Kametani*, C. Tarantini, E. Hellstrom

*Email: kametani@asc.magnet.fsu.edu

*Applied Superconductivity Center, National High Magnetic Field Laboratory, Florida State University, Tallahassee, FL, 32310, USA*

**Abstract**

We propose a targeted attack to make the Fe-based superconductor (FBS) $AEFe_2As_2$ (122) (AE = alkaline earth, here Ba or Sr) into a much cheaper, higher stability margin, round, twisted and effectively isotropic multifilament conductor suitable for 16-20 T dipole magnet construction. Compared to $Nb_3Sn$ which is now in its last asymptote of innovation, 122 is still at the early stage of development. This white paper urges to greatly accelerate progress by providing a fundamental understanding of what controls the transport critical current density, $J_c$, of 122 polycrystals and then fabricating it into wire form. High grain-to-grain $J_c$ is the one key property not yet fully demonstrated in this new superconductor and this is the key breakthrough that would make 122 practical. Whereas the intragrain (i.e. vortex pinning) $J_c$ is already high enough to replace $Nb_3Sn$, our recent work strongly suggests that unpredictable superconducting connectivity at grain boundaries (GBs) is the critical problem that limits the long-range $J_c$. The most important questions to increase the long-range $J_c$ are: What degrades the GB connectivity in 122? How can we overcome this GB degradation yielding high $J_c$? Can we incorporate the processes that increase $J_c$ into powder-in-tube fabrication yielding inexpensive, high-performance wires? These questions are crucial for 122 to become the affordable 16-20 T magnet technology that the Particle Physics Project Prioritization Panel (P5) and the US Magnet Development Program (MDP) strongly demand. In the next 10 years, we propose to address the following key issues so that FBS's potential can be fully evaluated and transformed into a magnet-ready conductor with $J_c$ >1,500 $A/mm^2$ at 4.2 K and 16-20 T: Define the most effective synthesis for high quality 122 powder to eliminate extrinsic current blockers at GBs; Comprehensively evaluate the true nature of GB superconductivity; Define the most effective wire design and fabrication route to transfer the effective synthesis into long-length multi-filamentary wire form. These thrusts are inter-related and must be addressed in series in a strategic manner with long term commitment. We strongly emphasize the need to evaluate the true intrinsic nature of GBs in the polycrystalline forms before focusing on wire fabrication.

# 1. Introduction

Future particle physics discoveries at the energy frontier rely on a next generation of high-energy colliders of order 100 TeV. US strategy has been set by the Particle Physics Project Prioritization Panel P5 [1], which has strongly supported a future high-energy proton-proton collider. Globally CERN is extensively investigating machine designs for a 100 TeV Future Circular Collider (FCC) [2]. There are also growing design activities for a Super proton-proton Collider (SppC) in China [3]. Such colliders require about twice the field possible from Nb-Ti, thus requiring new superconducting accelerator magnets of about 16 T. $Nb_3Sn$ is the logical superconductor but the P5 goals are for transformational high field magnet R&D that will deliver substantial magnet performance increases with a simultaneous dramatic cost reduction: these are great challenges for $Nb_3Sn$.

In response to these P5 goals, the US Magnet Development Program (MDP) defined four primary responses [4], one of which set ambitious goals to significantly increase performance while simultaneously reducing cost for 16-20 T dipole magnet construction. According to the recent FCC design studies, the cost of superconducting wire is a major portion of the total cost for a gigantic facility such as LHC or FCC. For example, the 16 T accelerator magnets in one FCC design will use 6,000-8,000 tons of $Nb_3Sn$ conductor, costing well over 10 B$ just for the conductor. A key problem is the high raw material price of Nb (~$420/kg for the required high purity Grade I Nb) [5]. The volumetric cost of HTS such as $REBa_2Cu_3O_{7-\delta}$ coated conductors (REBCO) or $Bi_2Sr_2CaCu_2O_x$ round wires (Bi-2212) is 6-10 times higher than $Nb_3Sn$ at present. CERN design studies indicate that 4 T HTS inserts for 16 T $Nb_3Sn$ dipoles will need 1,500-2,000 tons of HTS and 4,000-6,000 tons of $Nb_3Sn$ [6,7] that together will cost 20-30 B$. Obviously, such costs are grave obstacles to FCC. Many believe that the P5 and MDP magnet cost reduction goals are unlikely just by optimizing magnet design. Interestingly, a recent Fermilab cost analysis showed that cost of a future collider could decrease by 66 to 75% if the cost of magnets goes down by 80% [8].

*Conductor cost consideration*

Given that planning substantial magnet cost reduction is critical for the next 10 years, we must seriously consider the conductor cost. Table 1 tabulates conductor properties and cost per volume (liter) of representative state-of-the-art LTS and HTS superconductor candidates for accelerator magnets. Also shown are volumetric costs of the 122 (($Ba_{0.6}K_{0.4})Fe_2As_2$) conductors that we propose. The advantage of HTS cuprates (REBCO and Bi-2212) in $J_c$ (20 T, 4.2 K) over $Nb_3Sn$ is clear, while the high $J_c$ of single crystal 112 FBS clearly shows its native potential as a high field superconductor. Very striking is the high cost of all present conductors except Nb-Ti. Bi-2212 and REBCO are around $100 k per liter. High REBCO cost

*Table 1 Cost and property comparison of candidate superconductors for accelerator magnets*

| Material | P factor[*] | Conductor cost per liter[†] | $T_c$ (K) | $\mu_0 H_{c2}$ (4.2 K) | Intragrain $J_c$ (20 T, 4.2 K)[‡] | Intergrain $J_c$ (20 T, 4.2 K)[‡] | Conductor $J_E$ (20 T, 4.2 K) |
|---|---|---|---|---|---|---|---|
| Nb-Ti[§] | 3 | ~$2,000 | 9.2 | 11 T | n/a | 0 A/mm$^2$ | 0 A/mm$^2$ |
| $Nb_3Sn$[§] | 6~7 | ~$15,000 | 18 | 26 T | n/a | ~400 A/mm$^2$ | ~280 A/mm$^2$ |
| Bi-2212[§] | ~10 | ~$100,000 | 85 | ~100 T | n/a | ~6,000 A/mm$^2$ | ~1,200 A/mm$^2$ |
| REBCO[§] | >>10 | ~$100,000 | 92 | >120 T | ~20,000 A/mm$^2$ | ~20,000 A/mm$^2$ | ~420 A/mm$^2$ |
| **122 FBS** | **~10[#]** | **< $4,000** | **39** | **~90 T** | **~6,000 A/mm$^2$** | **~100-1000 A/mm$^2$** | **~15-200 A/mm$^2$** |

[*] *Production factor (P factor): the ratio of final product cost to raw material cost.*
[†] *Small high field magnets being made at the NHMFL typically use about one liter of conductor.*
[‡] *Intragrain and intergrain $J_c$'s are $J_c$ of a single crystal and $J_c$ across the grain boundaries, respectively.*
[§] *Property values are quoted for production grade tapes and wires.*
[#] *P factor of 122 is assumed to be the same as Bi-2212*

derives from present low yields and large capital costs of the complex and delicate fabrication process. For Bi-2212, ~60-65% of the whole wire cross section is Ag, which makes Bi-2212 wires expensive. Also we can see that the Nb price ($Nb_3Sn$ requires the grade I high purity Nb at ~ $420/kg, not the low purity industrial grade which costs ~$75/kg) significantly drives up the cost of Nb-Ti and $Nb_3Sn$. Nb-Ti is completely mature and mass-produced, making its production cost factor (P factor: the ratio of final product cost to the raw material cost) ~3. It costs ~$2k/liter. The price of $Nb_3Sn$ for HiLumi is ~$15k/liter with its present P factor of 6-7. Strikingly, the raw material cost of 122 is just $400/liter based on our recent purchase prices for small quantities of K-122 raw materials as lab supplies, making us believe that raw material costs can be decreased significantly for larger conductor productions. Even for a P factor = 10, 122 can be ~$4k/liter as a conductor, about one quarter the cost of present $Nb_3Sn$. What 122 still needs now is to increase the intergrain $J_c$ (~100-1000 A/mm$^2$) as close to the 122-FBS single crystal $J_c$ (~6,000 A/mm$^2$) as possible.

*122 FBS as the candidate for a high field magnet conductor*

FBS is the new class of HTS whose first compound $LaFeAsO_{0.89}F_{0.11}$ was discovered to exhibit superconductivity at $T_c$ ~26 K in 2008 [9]. Since then, several different families of FBS compounds have been discovered with $T_c$ up to 55 K and estimated $H_{c2}$ exceeding 100 T. With higher $T_c$ and $H_{c2}$, 4.2 K and 16-20 T use becomes foreseeable, especially in $AEFe_2As_2$ (122; AE = Ba, Sr or Ca) because of its low $H_{c2}$ anisotropy. Figure 1 compares the temperature dependence of $H_{c2}$ (or $H_{irr}$) of $Ba(Fe_{0.9}Co_{0.1})_2As_2$ (Co-Ba122) and $Ba_{0.6}K_{0.4}Fe_2As_2$ (K-Ba122) to other superconductor candidates for accelerator magnets. 122 compounds have the required high $T_c$ and $H_{c2}$, to clear the first hurdle to becoming the accelerator conductor of the future. The goal of the proposed work stated in this white paper is to provide the scientific and technological foundation for making 122 into a conductor technology with the possibility of providing a much cheaper, higher stability margin, multifilamentary twisted conductor with the flexible architectures of Nb-Ti and $Nb_3Sn$.

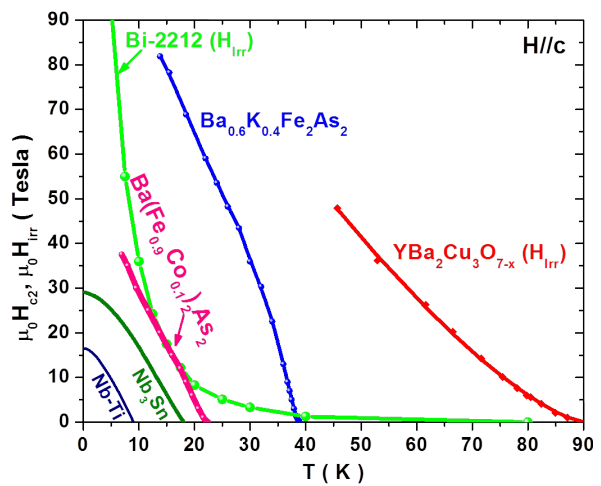

Figure 1 Temperature dependence of upper critical field ($H_{c2}$) or irreversibility field ($H_{irr}$) of various superconductor candidates for accelerator magnets. Note the 3-4 times higher values for the cuprate and Fe-based superconductors as compared to $Nb_3Sn$.

Also, the doping diversity of 122 allows us to place dopants into any of the atomic sites in the crystal structure. Co or Ni substitution for Fe adds electrons while K substitution for Ba (Sr) adds holes, both of which induce superconductivity. As with the cuprates, hole doping generates higher $T_c$ and $H_{c2}$. Chemical pressure in the crystal exerted by substituting As with isovalent elements (e.g. P) also makes 122 superconducting. This doping flexibility will be a great tool to assess the correlations of carrier density, $H_{irr}$, and $T_c$ to current transport in polycrystals. If $J_c$ can be raised in polycrystalline 122 by one order of magnitude towards its single crystal value listed in Table 1, then 122 could be truly transformational for high-field magnet technology due to the extra stability margin provided by a $T_c$ in the >30 K range.

*Challenges of 122 FBS to be a high field magnet conductor*

There are many reports about the 122 wire and/or tape fabrication, some of which reported $J_c$ > 1000 A/mm$^2$ at 4.2 K, 10 T [10]. However, the major challenge to boost $J_c$ to meet the Future Circular Collider, FCC, target ($J_c$ = 1500 A/mm$^2$ at 4.2 K, 16 T) is increasing 122's intergrain $J_c$ ($J_c$ across the grain boundaries (GBs)). Currently 122's intergrain $J_c$ is suppressed by degraded superconducting connectivity across GBs. In the past 3 years, groups in China and Japan have been more focused on wire fabrication without clear understanding of processing and of what determines $J_c$. $J_c$ of their short monofilament flat tapes has been increased by what appears to be a trial-and-error approach that unfortunately hasn't provided useful scientific guidelines about what limits $J_c$ or how to systematically improve 122's performance. As a result, they have not been successful increasing $J_c$ further, nor in reproducing their short-sample high $J_c$ in long lengths and/or in multifilamentary wire. This demonstrates that there are still multiple extrinsic current blockers in any 122 and that the correlations between the synthesis processing, composition and extent of extrinsic blockers are still poorly understood. Thus for 122, we must begin by focusing on the fundamental issues related to GB connectivity before spending resources on wire fabrication. In the next 10 years, the following questions to increase the intergrain $J_c$ should be addressed:

1. What degrades the GB connectivity in 122?
2. How can we overcome this GB degradation yielding high $J_c$?
3. Can we incorporate the processes that increase $J_c$ into powder-in-tube fabrication yielding inexpensive, high-performance wires?
4. How can we implement those processes into long-length, multi-filamentary conductor technology?

The intrinsic $J_c$ characteristic at the GBs is still unknown; however, our recent studies strongly suggest that the GB connectivity in state-of-the-art 122 is still significantly degraded by extrinsic current blockers such as impurity phases and nanocracks that must be eliminated to determine the intrinsic limitation of $J_c$ across the GBs. Understanding these $J_c$ limitations and improving 122's performance requires full, fundamental knowledge of the correlations between synthesis chemistry, intragrain pinning, and intergrain connectivity in polycrystalline 122.

## 2. Goals

Our first goal should be to further understand and eliminate all extrinsic current blockers so that we can evaluate the true intrinsic GB connectivity in polycrystalline 122. Key to achieving this goal is synthesizing and fabricating polycrystalline 122 samples with very clean, fully connected GBs. Even in the samples made using the present clean synthesis methods, we have not fully achieved strong GB connectivity because there are still extrinsic current blockers such as nanocracks or impurity phases that form due to hidden variables in the chemical synthesis, which we need to identify and understand to fully address the intrinsic GB superconducting properties. This requires further optimizing the processing and improving the GB chemistry followed by extensive microstructural and electromagnetic characterization. Then after evaluating the true intrinsic nature of 122 GBs, our goal is to gradually transition to investigate wire fabrication such as the Powder-In-Tube with multi-filamentary architecture. The outer sheath configuration will be investigated and adopted for the long-length wire forms.

## 3. Present status on 122 FBS

### 3.1  Development of a clean synthesis protocol

We have focused on synthesizing clean K-doped BaFe$_2$As$_2$, which provides the foundation to evaluate the true intrinsic current limiting factors in 122. We correlated the nanostructure to superconductivity performance as part of addressing the synthesis issues. We use hot isostatic pressing (HIP) during synthesis to eliminate porosity and macroscopic cracks that were common current blocking obstacles in early 122 and other FBS polycrystals [11–13]. However, even with HIP processing there are still many unknown variables in the 122 synthesis.

To synthesize cleaner 122, we reduced the $O_2$ and $H_2O$ content in the synthesis environment using a new high-performance glove box and a new high-energy planetary ball mill with sealed milling jars. The new glovebox reduces the $O_2$ and $H_2O$ concentrations below 0.005 ppm and 0.05 ppm, respectively, which are significantly lower than in our old glove box, which had $O_2$ and $H_2O$ concentrations of 100 and 10 ppm, respectively.

Synthesizing samples in the new glove box successfully eliminated almost all oxygen segregation at GBs. In the samples made in the old glove box, a network of Ba-O formed at the GBs that impeded the intergrain connectivity [14]. However, samples made in the new glove box exhibited strong K segregation at the GBs (Figure 2). Surprisingly, despite the K segregation, $J_c$ increased by as much as 70 % in the cleaner samples. This indicated that all possible oxygen sources need to be eliminated to fully explore the interrelations between chemical reactions, such as this K segregation.

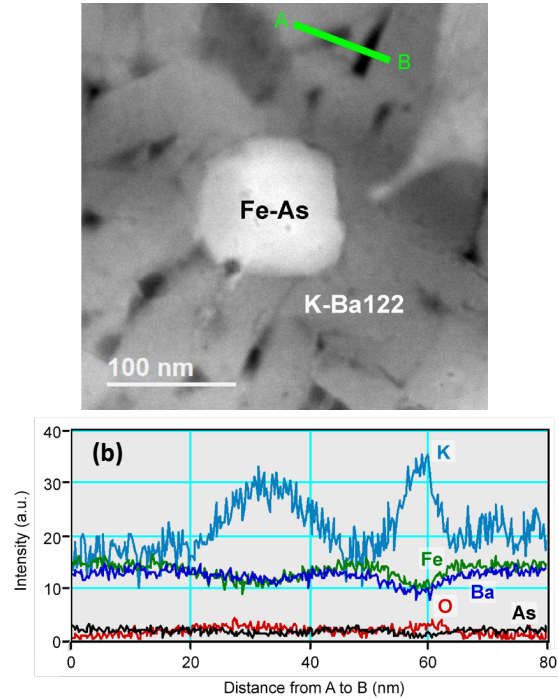

Figure 2 (a) ADF-STEM image of a representative GB in a sample prepared in the new glove box with a long milling and a short heat treatment. (b) Compositional variations derived from the EELS line scan along the green line from A to B shown in (a).

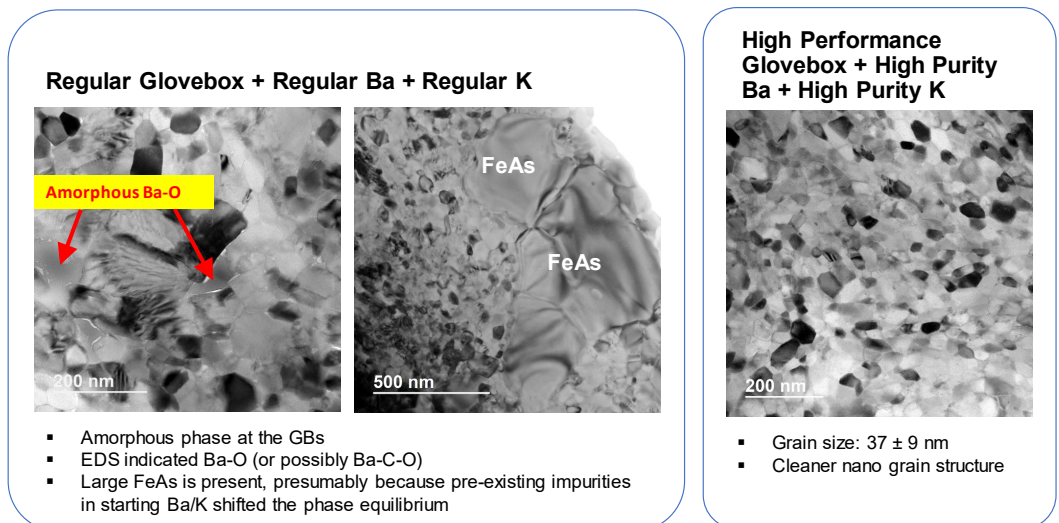

Figure 3 TEM characterization of 122 samples obtained by different processing condition. On the left, a sample (shown at two magnifications) prepared with standard-purity Ba and K in the old regular glove box showing amorphous Ba-O at GBs and a large amount of FeAs secondary phase. The image on the right, which is a sample prepared in the new, high-performance glove box using high-purity Ba and K, does not show the amorphous Ba-O and FeAs.

We also switched to high-purity Ba and K (>99.95 % purity). This improved intergrain superconducting connectivity by effectively eliminating the FeAs wetting phase at GBs present in samples made with lower purity Ba and K in the old glove box. A sample made with lower purity Ba and K in the old glove box is shown in the left 2 images in Figure 3. Analytical scanning TEM of samples made with the improved synthesis route found no undesirable GB segregation and FeAs was significantly reduced (right image in Figure 3). Furthermore, K segregation at GBs was no longer observed.

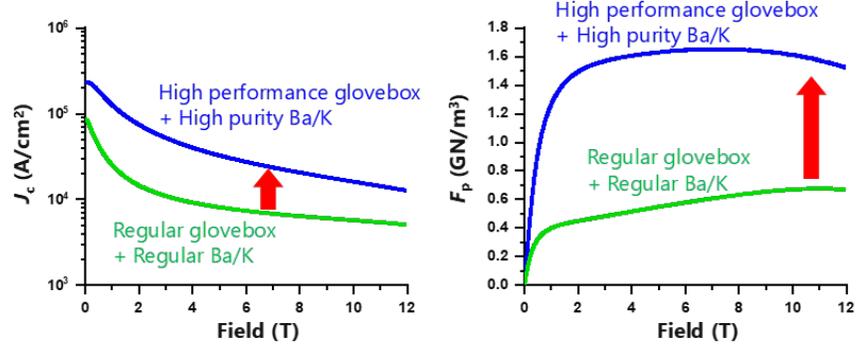

*Figure 4 Comparison of (left) $J_c$ and (right) $F_p$ field dependences for the best sample prepared with regular-purity Ba and K in the old glove box and the best sample prepared in the new glove box with high-purity Ba and K. Our new clean synthesis more than doubled $J_c$ and $F_p$.*

After establishing our clean synthesis protocol, we re-examined the optimum heat treatment temperatures. The synthesis starts with a 1st ball milling of the starting elements, a 1st heat treatment (1st HT) in a HIP, a 2nd ball milling followed by a 2nd HT in a HIP. We began by studying changes in the 1st HT temperature. Starting from 600 °C, based on Weiss *et al*. [11,15], we used 600, 675, 750, and 825 °C for the 1st HT and kept the 2nd HT at 600 °C. Figure 4 shows that optimizing the 1st HT temperature essentially doubled $J_c$(4.2 K) to $2.3\times10^5$ A/cm$^2$ at self-field and $1.6\times10^4$ A/cm$^2$ at 10 T [16].

Our systematic study of synthesis routes strongly suggests that a clean synthesis environment, high purity Ba and K, and a proper 1st HT temperature are crucial to enhance the superconducting connectivity in the K-Ba122 polycrystals. However, our best $J_c$ is still only ~10-20% of single-crystal $J_c$ of the same composition [16], indicating that even in these clean samples, GBs are still degrading connectivity.

### 3.2 Understanding the correlation between synthesis and superconductivity

Since no correlation was found between the magnetization $T_c$ onset ($T_{c,\text{mag}}$ evaluated by SQUID measurements) and $J_c$ in the clean samples described above, we systematically investigated the effects of heat treatment (by varying either the 1st or 2nd HT temperature from 600 up to 825 °C) on bulk superconducting properties of 122 using specific heat characterization [17]. The advantage of this technique is that it is not affected by GB connectivity and it offers an overall investigation of the sample properties detecting differences on the scale of ξ, the coherence length. In particular, we evaluated the $T_c$-distributions of different samples, which have a sharp peak at the transition (Figure 5), revealing subtle but important differences between samples. We found that at zero-field the peak position of the $T_c$-distribution is shifted to higher temperature after performing the 1st and 2nd HT at 750 and 600 °C, respectively (which produced the best $J_c$-performing sample), when compared to a sample prepared at 750 + 750 °C (worst sample). On the best sample we also

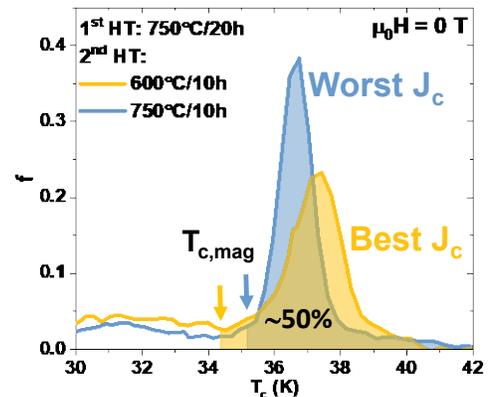

*Figure 5 $T_c$-distributions of the best and worst samples in terms of $J_c$ investigated in the HT optimization study. The shaded areas represent the fraction (f) of the distribution above the magnetization $T_c$ onset, $T_{c,mag}$.*

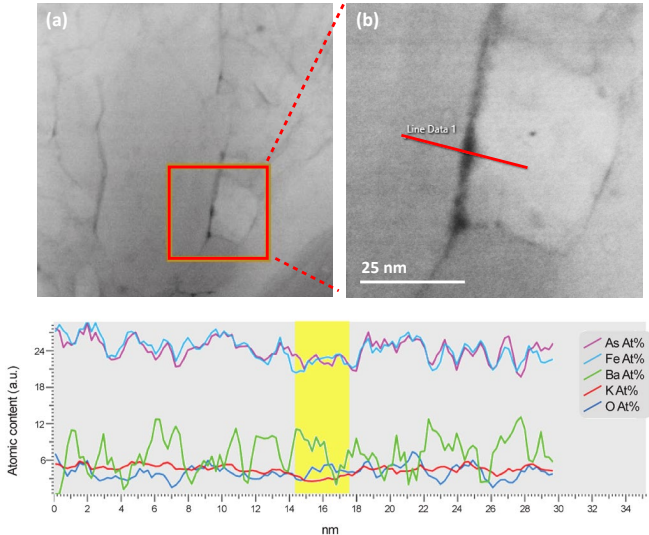

*Figure 6 (a,b) GBs in samples made with our clean synthesis protocol (at two magnifications) do not show chemical degradation, but the dark trace of GBs indicates a lower material density at the GB due to GB nanocracks. (c) Energy Dispersive X-ray (EDS) along the line in (b) showed there is almost no chemical variation across the GB.*

observed the weakest field dependence of the $T_c$-distribution, implying a higher $H_{c2}$, which is an advantage for the low-temperature performance. Interestingly, we also observed a clear difference between the magnetization and specific heat $T_c$ transition (Figure 5) with $T_{c,\mathrm{mag}}$ being located well below the peak of the $T_c$-distribution. We ascribed this difference as being likely caused by a combination of the small grain size (grains are much smaller than the penetration depth) and extrinsic connectivity effects. This is important because it means that magnetization is unable to probe a significant part of the sample (~50%) that is crucial for the $J_c$ performance. Since 50% is a quite large fraction compared to the experimental values for other materials, such as $Nb_3Sn$, and to the theoretical percolation threshold (a percolative path is necessary to induce shielding in magnetization measurements), we think that extrinsic factors such as nanocracks still play a role in the connectivity degradation. This was confirmed by TEM and EDS (energy dispersive X-ray) analyses (Figure 6), which indeed revealed no chemical degradation at the GB but a dark contrast indicating a significantly decreased material density that could be due to a nanocrack.

To better understand the particular effect of the heat treatment on the connectivity and pinning properties, we also compared the low- and high-field $J_c$ to the peak position of the $T_c$-distribution (Figure 7). The low-field $J_c$ (top panel) is expected to be directly related to $T_c$ : if this does not occur, it means that connectivity is varying. On the other end, the high-field $J_c$ (bottom panel) can also be affected by the pinning performance. What we found is that increasing the 1st HT temperature up to 750 °C does not degrade the connectivity and has moderate effect at higher temperature, whereas any increase of the 2nd HT temperature is strongly detrimental for the connectivity. The high-field behavior suggests no significant changes in the pinning efficiency in samples with different 1st HT temperatures, whereas increasing the 2nd HT temperature causes a reduction of the pinning efficiency on top of the connectivity loss.

These results suggest a possible change in the synthesis route to improve $J_c$. Since the best bulk properties (highest $T_{c,peak}$ and highest $H_{c2}$), connectivity, and $J_c$ are found in the sample heat treated at the lowest tested 2nd HT temperature, lowering the 2nd HT temperature even further may improve the phase properties. In addition, decreasing the cooling rate in the HIP may eliminate possible nanocracks caused by thermal stress.

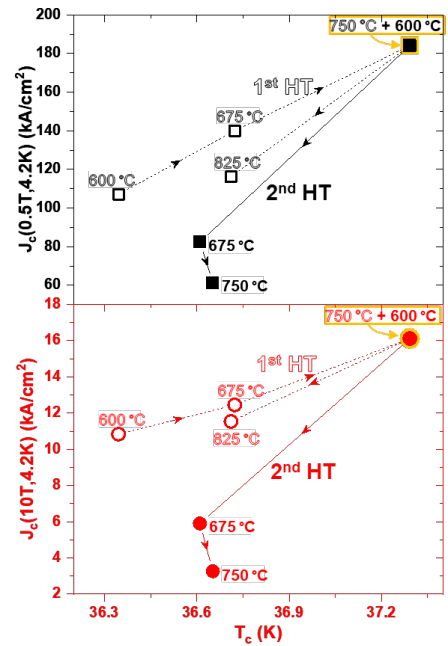

*Figure 7 Low and high-field $J_c$ at 4.2 K as a function of the peak position of the $T_c$-distribution (see Figure 6). The arrows between points indicate the effect of increasing the HT temperature of the 1st (open symbols) or the 2nd (full symbols) HT, respectively. Data from [17].*

## 3.3 Many GBs in 1000 A/mm$^2$ $J_c$ tapes are still contaminated by impurity phases

Polycrystalline 122 has made progress toward the FCC target ($J_c$ = 1500 A/mm$^2$ at 4.2 K, 16 T) [18,19]. A Chinese-Japanese collaboration led by Prof. Ma reported $J_c$ (15 T) of 1000-1100 A/mm$^2$ in a polycrystalline flat tape [20,21]. Tamegai *et al*. recently reported $J_c$ (5 T) of 600 A/mm$^2$ in round wires [22], which is preferred by magnet designers and builders. However, it is important to point out that these advances relied heavily on Edisonian trial-and-error during their wire fabrication processing without full fundamental understanding of the GB connectivity. We performed STEM/TEM analysis on some Ma's high $J_c$ tapes and surprisingly. observed significant amounts of secondary phases segregated at GBs even in the best $J_c$ tape (Figure 8). This strongly suggests that $J_c$ optimization on the wires and tapes has been done without cleaning the GBs and without fully understanding what causes the dirty GBs. Furthermore, the supercurrent paths and effective cross section for current flow in their tapes are unknown.

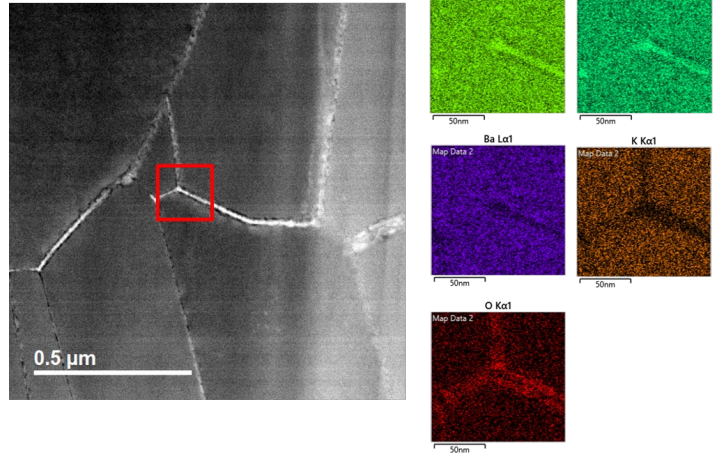

*Figure 8 Chemically degraded GBs found in a tape with a $J_c$(4.2 K, 10 T) = 10$^5$ A/cm$^2$. The red box in the left figure is a GB triple point that is analyzed in the right figures. These results show that many GBs are still compromised by FeAs and Ba-O.*

Nevertheless, $J_c$ (15T) of 1000-1100 A/mm$^2$ in their samples with many contaminated GBs shows 122's potential as a conductor technology if the GBs are fully connected by eliminating all extrinsic blockers. There is no concern about the intragrain $J_c$ in bulk samples because inherent strong vortex pinning was found in single crystals [23]. But understanding what causes the bottle neck degrading connectivity in 122 is still lacking. Is it intrinsic GB blockage, extrinsic GB blockage, and/or intrinsic $J_c$ anisotropy? Over the next 10 years, we must fully address this question from all angles of chemistry, nanostructure, and material science to determine the feasibility of 122 and then, if the GB issues can be resolved by powder processing and incorporated in wire fabrication, we will move onto establishing the next high-field conductor technology using 122.

## 4. Proposed plans

Since the full potential of 122 as a high field conductor is not fully explored yet, research on and development of 122 should be carried out step-by-step from the material fundamentals to conductor fabrication. The short-term goal over the next 5 years should be to further understand and eliminate all extrinsic current blockers so that we can evaluate the true intrinsic GB connectivity in polycrystalline 122. This step is very crucial because 122's feasibility as a PIT wire depends on whether the intrinsic GB connectivity issues exist. Even in the improved samples, strong GB connectivity is not fully achieved because there are still extrinsic current blockers such as nanocracks or impurity phases that form due to hidden variables in the synthesis chemistry that prevent us from fully addressing the intrinsic GB superconducting properties. The root causes of the low, unpredictable GB connectivity need to be fully investigated and resolved by further optimizing the processing and improving the GB chemistry followed by extensive microstructural and electromagnetic characterization. We focus on wire fabrication issues after the root causes of GB connectivity issues are addressed.

Our work plan is the following;

1. **Define the most effective synthesis for high quality polycrystalline 122 with strong intergrain connectivity**: Our clean synthesis method reveals that our understanding of the 122 synthesis reaction is still incomplete. The 122 synthesis consists of multiple high energy millings and heat treatments in a hot isostatic press. We need to fully understand the correlations between the starting composition, the multi-cycle milling and heat treatment, the synthesis environment, such as the external pressure that causes intimate contact with the crucible material, intragrain flux pinning, and intergrain superconducting connectivity. Deconvoluting the effects that these entangled synthesis parameters have on $J_c$ requires very clean synthesis protocols to make nearly 100% dense, polycrystalline 122 without extrinsic current blockers at the GBs.
2. **Relationship between micro/nanostructure and intra/intergrain properties**: What are the correlations between the micro/nanostructural changes due to the processing and the resulting intragrain/intergrain superconductivity in 122? We must fully address such a central question. The GB superconducting properties in 122 can influence intergrain transport and flux pinning, both of which affect the overall $J_c$. We need to evaluate the GB superconductivity by combining various electromagnetic measurements and microstructural analyses. This fundamental understanding of intragrain and intergrain superconductivity will help to determine the best synthesis route for 122, providing the key foundation for high-$J_c$ wire fabrication.
3. **Define the most effective wire design and fabrication route**: After we have increased $J_C$ in 122, we will turn our attention to fabricating high $J_c$ 122 wires. This requires proper selection of the material for the innermost sheath to avoid poisoning the 122. We will determine the most effective combination of drawing, groove rolling and swaging to produce uniform 122 filaments, and will fully optimize the heat treatment protocols. To achieve high-$J_c$ 122 wire, we must address the correlations among $J_c$, powder processing, wire design, wire fabrication, and heat treatment.

## 4.1 Define the most effective synthesis for high quality polycrystalline 122 FBS

One of the important challenges for making high-$J_c$ 122 is controlling the GB nanostructures. Currently the GB connectivity is not fully under control even in our best bulks, nor in other groups' best wires and tapes. There are different kinds of extrinsic current blockers at GBs including FeAs, Ba oxide, and/or nanocracks. Although we have reduced some of these, it is still largely unknown what synthesis parameter plays a role to effectively remove each of them. Thus, it is central for the 122 research to fully investigate the synthesis protocol from various perspectives.

*Further optimization of 122 composition*

The best chemical composition of 122 powder to attain strong intergrain connectivity and high $J_c$ in bulk samples may not necessarily be the exact stoichiometry of 122 single crystals that are optimally-doped to maximize $T_C$. As an example, the Bi-2212 composition with the highest $J_c$ is off-stoichiometry [24]. We plan to optimize the chemical composition of 122 to maximize $J_c$, intragrain pinning, and GB connectivity. It is still unclear whether the starting composition is preserved in the final intragrain composition of 122 after the final heat treatment. We performed compositional analysis using TEM on individual 122 grains in various 122 bulks and found that the average intragrain composition became off-stoichiometry relative to the starting composition as the 2$^{nd}$ HT temperature increases, resulting in monotonic reduction of $J_c$ (Table 2). The Ba/K ratio stays close to the optimum (Ba : K = 6 : 4) despite very minor K loss. However, the intragrain composition obviously became Fe-rich and/or As-rich, potentially degrading the intragrain superconducting properties. We need to fully understand the mechanism of the intragrain off-stoichiometry so that we can find the optimum starting composition to maximize $J_c$.

Table 2 Comparison of $J_c$ and average intergrain composition after 2nd HT by HIP. Note that the 1st milling, 1st HIP HT and 2nd milling parameters are identical. As the 2nd HT temperature increases, the average intragrain composition becomes off-stoichiometry.

| 2nd HT temperature | $J_c$ (4.2 K, 5 T) | Average intragrain composition | Ba/K ratio | Note |
|---|---|---|---|---|
| 600 °C, 10 hrs | $3.1 \times 10^4$ A/cm$^2$ | $Ba_{0.61}K_{0.39}Fe_{1.98}As_{2.34}$ | 6.06 : 3.94 | 18.5 % As-rich |
| 675 °C, 10 hrs | $1.3 \times 10^4$ A/cm$^3$ | $Ba_{0.64}K_{0.36}Fe_{2.42}As_{2.26}$ | 6.39 : 3.61 | 22.3 % Fe-rich, 12.7 % As-rich |
| 750 °C, 10 hrs | $0.9 \times 10^4$ A/cm$^4$ | $Ba_{0.63}K_{0.37}Fe_{2.47}As_{2.27}$ | 6.30 : 3.70 | 23.5 % Fe-rich, 13.6 % As-rich |

Another interesting observation is seen in Figure 9, which compares $J_c$ of our best bulk and Tamegai's round wire [20]. The target composition of both samples is the same, $(Ba_{0.6}K_{0.4})Fe_2As_2$, but the resulting $J_c$ behaviors are rather different. Because the GB connectivity in our bulks is slightly better than Tamegai's, we achieve a low-field bulk $J_c$ ~10 % higher than Tamegai's. However, a crossover occurs at 2 T and $J_c$ of our sample drops more quickly in high fields, indicating that our samples might lose intragrain pinning during the synthesis or conversely that intragrain pinning and/or $H_{Irr}$ is enhanced during their wire fabrication.

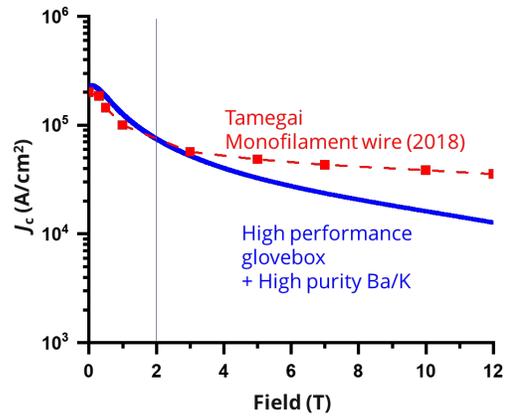

Figure 9 $J_c$(4.2 K) comparison between our best bulk and Tamegai's best round wire [20]. The target composition of both samples is $(Ba_{0.6}K_{0.4})Fe_2As_2$.

### Hidden variables in high pressure synthesis

TEM analysis of our bulk samples showed evidence of decreased density at GBs suggesting the presence of nanocracks (Figure 6), even in nearly 100% dense HIPped samples. The nanocracks may form because the pressure is suddenly released at the end of the HIP HT cycle, which rapidly cools the sample. We will investigate the physical effects of sudden pressure release during the synthesis, evaluating how the performance changes by varying the applied pressure during the synthesis processing and/or by changing the pressure-release/cool-down procedure.

Although HIP is very useful to fully densify bulk samples from the powder state [15], the high pressure forces the 122 and crucible material together in intimate contact, which may enhance any reactions that occur at the Nb/122 interface. Reactions between the 122 and Nb may be hidden synthesis variables. Interestingly, we recently observed some slight reactions between 122 and the Nb crucible material when synthesizing 122 bulks; however, the reactions' impact on the overall stoichiometry and superconductivity is not yet known. Nb may also diffuse along 122 GBs modifying the GB chemistry. Wires made by other groups generally have 122 in contact with an Ag inner sheath, which we think is one of the most inert sheath materials towards 122 (although the As-Ag phase diagram indicates that Ag may dissolve a small amount of As during the heat treatment). We need to determine the most inert crucible material and the most effective starting 122 composition to control the final composition.

### Optimization of milling and heat treatment

Our specific heat measurements have begun to show how the 1st and 2nd HTs affect $J_c$ [17]. But it is not yet clear how the milling step in the synthesis affects the phase uniformity and intergrain connectivity. Previous work on Co-doped Ba122 indicated that varying the milling energy density before the 1st HT affects $T_c$, $H_{Irr}$ and $H_{c2}$ [25,26]. We are conducting the milling experiments on K-doped Ba122 that shows that both the milling energy density for the 1st milling and 1st HT temperature significantly affect the final $J_c$, after the 2nd HT. Interestingly we found that, even maintaining the same 2nd milling and 2nd HT, the K doping level and amount of FeAs impurity phase widely varies, demonstrating that the effects of the 1st milling and 1st HT remain in the later stage of processing. This experiment also shows that the milling energy density influences $T_c$ before the 2nd HT, indicating the chemical homogeneity, reaction completeness, and/or grain size of the 122 phase depends on the milling energy density. Clearly, we still need to fully understand the correlation between the intragrain/GB composition, $T_c$, $H_{c2}$ ($H_{Irr}$) and intergrain $J_c$ with variations in ball milling energy density and HT temperature.

*What is the best dopant?*

We recently started to evaluate doping elements other than K. We have been using K as the doping element because $(Ba_{0.6}K_{0.4})Fe_2As_2$ single crystals show the highest $T_c$ among the Ba122 system. However, recent studies by Tamegai *et al*. suggest that higher $J_c$ in round wires can be obtained with Na-doped Ba122, despite its lower $T_c$ with respect to K-Ba122 [22]. We plan to improve $J_c$ by finding the most effective 122 dopant. $J_c$ comparison of clean K- and Na-122 bulks might give us new insights about differences of intragrain properties as well as intergrain connectivity related to the different chemical reactions.

## 4.2 Relationship between micro/nanostructure and intra/intergrain properties

### 4.2.1 How to evaluate the 122 intra- and intergrain superconducting properties

By combining magnetization and specific heat measurements to evaluate the superconducting transitions and magnetic $J_c$, our previous studies on bulk samples demonstrated a lack of correlation between the magnetization $T_c$ (determined by DC susceptibility) and the $J_c$ performance [7]. Specific heat measurements allowed us to understand that this discrepancy is due to a combination of grain-size effects and variation in connectivity when different heat treatments are performed. We need to develop the characterization schemes to separate the effects caused by intergrain connectivity from intragrain pinning density variations, which cannot be easily identified by VSM alone.

*Transport characterizations*

The easiest technique to determine variation in the bulk superconducting properties and in connectivity is in-field low-current transport characterization*.* The resistive field/temperature dependences on bulk samples provides a direct evaluation of $H_{c2}$(T) and $H_{Irr}$(T). $H_{c2}$(T) provides information about the mostly intragrain disorder that determines the high-field performance limit. $H_{irr}$(T) carries information about both pinning and connectivity and, together with the evaluation of the normal state resistivity $\rho_n$, could provide valuable hints about the phase connectivity and/or presence of current-blocking secondary phases (e.g. $\rho_n$ was a valuable tool to determine connectivity issues caused by secondary phases at GBs in $MgB_2$ [27,28]). *High-current transport characterization* may be needed on some occasions to evaluate the in-field $J_c$ of monofilamentary short wire samples in our 15 T magnet. Such occasional fabrication of monofilament wires will enable us to change the 122 grain alignment and GB morphology. In fact, even monofilament round wire can develop some grain alignment.

*Magnetic characterization*

We will also consider other magnetic characterization techniques, such as remnant magnetization measurements. Remnant magnetization has already been employed in FBS materials and it was instrumental in identifying the orders of magnitude difference in the local (intragrain) and the global (intergrain) $J_c$ caused by connectivity issues at GBs in the early non-optimized 1111 (REFeAsO, RE: Rare Earth) samples [29,30]. Since remnant magnetization characterization allows estimating both intragrain and intergrain $J_c$ in bulk samples, as well as the intragrain $J_c$ in powder samples, we plan to employ it at different stages during the synthesis to monitor how each processing stage affects the intra- and intergrain superconductivity.

### 4.2.2 Micro-, nano- and atomic-structure characterization

Since the characteristic diameter of the vortex is only ~5 nm, $J_c$ of 122 is affected by overall microstructure and GB nanostructure, which strongly depend on various synthesis parameters. An aim of this white paper is to address the true intrinsic properties of GBs in polycrystalline 122 with very clean, well-connected GBs. Thus, we need to continue actively investigating the micro-/nanostructures of 122 grains and GBs using various kinds of electron microscopes. The scale of current blocking factors in the superconducting 122 ranges from micron down to angstrom size scales including micro- to nanoscopic size cracks and micro- to nanochemical compositional variations at the GBs. So, we need a comprehensive structural description on many different length scales. As an example of such analysis, Figure 10 presents the XRD results obtained at FSU, revealing that the $1^{st}$ milling and HT condition strongly influences the amount of FeAs phase in the final samples after the $2^{nd}$ HT. It is also important to perform *3D atom probe tomography* (APT) on our key samples to investigate possible element segregation at GB [31], because segregation at the GBs blocking the physical connection of 122 grains is still possible, maybe causing the GB nanocracks.

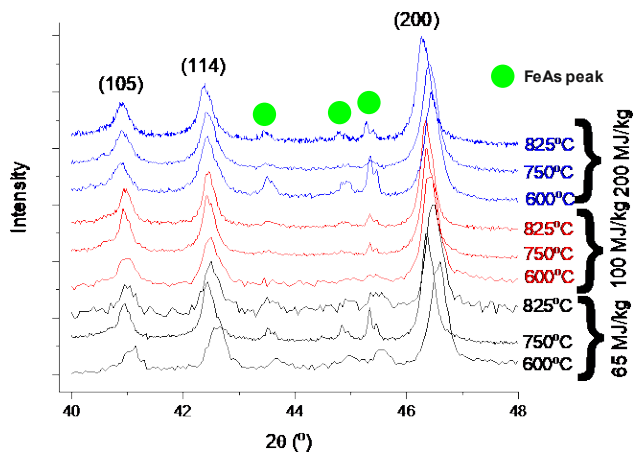

Figure 10 The XRD patterns of 122 FBS bulks after the $2^{nd}$ HT. Each sample was prepared with different $1^{st}$ milling energies and $1^{st}$ HT temperatures as indicated, but maintained the same $2^{nd}$ milling and HT condition.

## 4.3 Define the most effective wire design and wire fabrication route

After we have increased $J_c$ in bulk samples to the levels needed for applications, we will begin to investigate making 122-FBS wire in earnest so we can truly evaluate how far 122 can go as a high-field superconducting wire technology. In addition to optimizing the synthesis and heat treatment process for the wires, we will address the underlying engineering and scientific issues of wire fabrication. Below we lay out the key issues/challenges for fabricating 122-FBS wires. Our goal is to find the most effective wire architecture design and fabrication route to fabricate high-$J_c$ wire. We will start with making a mono filament architecture, because it is the simplest to fabricate and the easiest to manage the metallurgy issues such as the filament discontinuity or sausaging during the powder-in-tube process. Multi-filament architecture will be investigated once the effective sheath design is established.

### 4.3.1 Sheath materials

A 122-FBS wire has to be HTed to improve the superconducting propertied and achieve a good connectivity, and chemical reactions that can occur during the HT determine applicable sheath materials. Figure 11 shows the schematic and an actual cross section of a monofilament 122 wire we fabricated. The 122 powder is encased in the inner Ag sheath that is inside a Cu outer sheath. Almost all 122 wires and tapes being made by other groups have the same sheath configuration[20,32]. As mentioned above, Ag is widely used in contact with 122 because there is little reaction between Ag and 122, but there is evidence of some reaction that we need to quantify. Generally Cu is used for the outer sheath because it cannot be in direct contact with 122 [33]. Drawbacks to using Cu in direct contact with Ag are that Cu can diffuse through the Ag during the HT poisoning the 122, and the Ag-Cu eutectic reaction limits HT temperatures to < 779 °C. We may need to consider adding a barrier layer, such as Nb or Ta, between the Ag and Cu, or replace Ag with a different inner sheath material, such as Nb or Ta. We could also replace all or part of the outer Cu sheath with stronger, harder materials such as stainless steel or Monel [21,34].

### 4.3.2 Wire final heat treatment

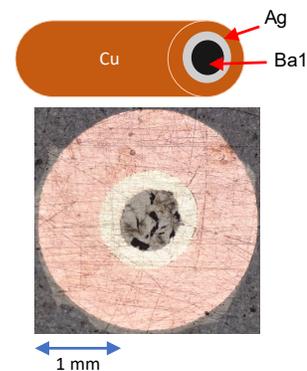

Fabricating wires may require additional or different synthesis processing than bulks. Referring to Figure 6, when we fabricated our first wire, we added a 3rd ball milling step to make the powder for the PIT wire, and a 3rd HT, heating the 122 PIT wire in the HIP. We call this Route 1. Although our recent round wires showed slightly higher $J_c$ than the bulk from which the powder was obtained even without optimizing the wire fabrication (Figure 11), it is possible that some fraction of the elements are lost during the multiple millings/HT cycles, so the 122 core in the wire is off-stoichiometry. To account for this loss, all groups including us, add 5-10 % extra of the volatile elements (K (Na) or As) to the starting mixture to compensate for this loss, although it is not very clear when or how those elements are lost. Nevertheless, the more milling/HT cycles we have, the greater the possibility of losing elements. To minimize these losses, we plan to investigate synthesis Route 2 for fabricating wires, where powder from the 2nd ball milling is used to fabricate the PIT wire and the 2nd HT is done on the PIT wire. Route 2 may be better to maintain the initial chemical composition in the final wire. We will investigate and optimize the Route 2 synthesis procedure for 122 wire. It is important to optimize the temperature and duration for the wire HT, which will be guided by the processing we will have developed for the high $J_c$ bulk samples.

*Figure 11 (Upper) Schematic illustration of our present wire configuration. (Lower) Representative cross section of our K-Ba122 monofilament wire.*

## 5. Conclusions

The 122 community rushed to fabricate wires without first solving the GB connectivity problem. As a results, there has been little improvement in wire $J_c$ performance, and $J_c$ reproducibility is poor in wires. This white paper outlines the research needed to understand what causes poor connectivity in 122 and how to improve the connectivity to achieve $J_C$ values mandated for practical applications. In short, this requires fully evaluating the GB superconducting connectivity of 122 by developing synthesis protocols to produce polycrystalline samples with very clean, fully connected GBs. The next step is fabricating wires from this high $J_c$ powder. This requires establishing the effective wire architecture and optimizing the wire fabrication and heat treatment protocols for the conductor technology. Our long-term goal is to fabricate high-performance 122 wires that will have the required performance at a significantly lower cost than present high-field conductors for the next generation of high-energy accelerator magnets recommended by P5 and MDP.

**REFERENCES**

[1] P5, Building for Discovery: Strategic Plan for U.S. Particle Physics in the Global Context, the Particle Physics Project Prioritization Panel (P5) Report: http://science.energy.gov/~/media/hep/hepap/pdf/May-2014/FINAL_P5_Report_Interactive_060214.pdf

[2] Benedikt M 2014 Future Circular Collider (FCC) Study, *May 3*

[3] Tang J, Berg J S, Chai W, Chen F, Chen N, Chou W, Dong H, Gao J, Han T, Leng Y, Li G, Gupta R, Li P, Li Z, Liu B, Liu Y, Lou X, Luo Q, Malamud E, Mao L, Palmer R B, Peng Q, Peng Y, Ruan M, Sabbi G, Su F, Su S, Stratakis D, Sun B, Wang M, Wang J, Wang L, Wang X, Wang Y, Wang Y, Xiao M, Xing Q, Xu Q, Xu H, Xu W, Witte H, Yan Y, Yang Y, Yang J, Yuan Y, Zhang B, Zhang Y, Zheng S, Zhu K, Zhu Z and Zou Y 2015 Concept for a Future Super Proton-Proton Collider *arXiv:1507.03224 [hep-ex, physics:physics]*

[4] MDP, The U.S. Magnet Development Program Plan: https://www.google.com/url?sa=t&rct=j&q=&esrc=s&source=web&cd=3&cad=rja&uact=8&ved=0ahUKEwi_j8e1i6fQAhVhjVQKHZI-DmgQFggrMAI&url=http%3A%2F%2Fwww2.lbl.gov%2FLBL-Programs%2Fatap%2FMagnetDevelopmentProgramPlan.pdf&usg=AFQjCNE_5dQydHgQdPN2r0ss-w4eKRHHQg&sig2=_RqcMNESvK_LLDsCj6pEzQ

[5] Cooley L and Pong I Cost drivers for very high energy p-p collider magnet conductors 20

[6] P. Fessia and L. Bottura, The CERN high field magnet programs, *International conference of high energy physics*, Aug. 03-10 2016, Chicago, IL

[7] Bottura L, Gourlay S A, Yamamoto A and Zlobin A V 2016 Superconducting Magnets for Particle Accelerators *IEEE Trans. Nucl. Sci.* **63** 751–76

[8] V. Shiltsev, Will there be energy frontier colliders after the LHC?, *International conference of high energy physics*, Aug. 03-10 2016, Chicago, IL

[9] Kamihara Y, Watanabe T, Hirano M and Hosono H 2008 Iron-Based Layered Superconductor La1-xFxFeAs (x = 0.05–0.12) with Tc = 26 K *J. Am. Chem. Soc.* **130** 3296–7

[10] Gao Z, Togano K, Zhang Y, Matsumoto A, Kikuchi A and Kumakura H 2017 High transport Jc in stainless steel/Ag-Sn double sheathed Ba122 tapes *Supercond. Sci. Technol.* **30** 095012

[11] Weiss J D, Jiang J, Polyanskii A A and Hellstrom E E 2013 Mechanochemical synthesis of pnictide compounds and superconducting $Ba_{0.6}K_{0.4}Fe_2As_2$ bulks with high critical current density *Supercond. Sci. Technol.* **26** 074003

[12] Weiss J 2015 *Synthesis and Characterization of Superconducting Ferropnictide Bulks and Wires* (Tallahassee, Florida: Florida State University)

[13] Kametani F, Li P, Abraimov D, Polyanskii A A, Yamamoto A, Jiang J, Hellstrom E E, Gurevich A, Larbalestier D C, Ren Z A, Yang J, Dong X L, Lu W and Zhao Z X Intergrain current flow in a randomly oriented polycrystalline $SmFeAsO_{0.85}$ oxypnictide *Appl. Phys. Lett.* **95** 142502

[14] Kametani F, Su Y-F, Collantes Y, Pak C, Tarantini C, Larbalestier D and Hellstrom E 2020 Chemically degraded grain boundaries in fine-grain Ba$_{0.6}$K$_{0.4}$Fe$_2$As$_2$ polycrystalline bulks *Appl. Phys. Express* **13** 113002

[15] Weiss J D, Tarantini C, Jiang J, Kametani F, Polyanskii A A, Larbalestier D C and Hellstrom E E 2012 High intergrain critical current density in fine-grain (Ba$_{0.6}$K$_{0.4}$)Fe$_2$As$_2$ wires and bulks *Nature Mater* **11** 682–5

[16] Pak C, Su Y F, Collantes Y, Tarantini C, Hellstrom E E, Larbalestier D C and Kametani F 2020 Synthesis routes to eliminate oxide impurity segregation and their influence on intergrain connectivity in K-doped BaFe$_2$As$_2$ polycrystalline bulks *Supercond. Sci. Technol.* **33** 084010

[17] Tarantini C, Pak C, Su Y-F, Hellstrom E E, Larbalestier D C and Kametani F 2021 Effect of heat treatments on superconducting properties and connectivity in K-doped BaFe$_2$As$_2$ *Scientific Reports* **11** 3143

[18] Ballarino A, Hopkins S C, Bordini B, Richter D, Tommasini D, Bottura L, Benedikt M, Sugano M, Ogitsu T, Kawashima S, Saito K, Fukumoto Y, Sakamoto H, Shimizu H, Pantsyrny V, Abdyukhanov I, Shlyakov M, Zernov S, Buta F, Senatore C, Shin I, Kim J, Lachmann J, Leineweber A, Pfeiffer S, Baumgartner T, Eisterer M, Bernardi J, Malagoli A, Braccini V, Vignolo M, Putti M and Ferdeghini C 2019 The CERN FCC Conductor Development Program: A Worldwide Effort for the Future Generation of High-Field Magnets *IEEE Transactions on Applied Superconductivity* **29** 6000709

[19] Tommasini D, Auchmann B, Bajas H, Bajko M, Ballarino A, Bellomo G, Benedikt M, Bermudez S I, Bordini B, Bottura L, Buzio M, Dhalle M, Durante M, de Rijk G, Fabbricatore P, Farinon S, Ferracin P, Gao P, Lackner F, Lorin C, Marinozzi V, Martinez T, Munilla J, Ogitsu T, Ortwein R, Perez J, Prioli M, Rifflet J-M, Rochepault E, Russenschuck S, Salmi T, Savary F, Schoerling D, Segreti M, Senatore C, Sorbi M, Stenvall A, Todesco E, Toral F, Verweij A P, Volpini G, Wessel S and Wolf F 2017 The 16 T Dipole Development Program for FCC *IEEE Trans. Appl. Supercond.* **27** 4000405

[20] Huang H, Yao C, Dong C, Zhang X, Wang D, Chen Z, Li J, Awaji S, Wen H, and Ma Y 2018 High transport current superconductivity in powder-in-tube Ba0.6K0.4Fe2As2 tapes at 27T *Supercond. Sci. Technol.* **31** 015017

[21] Gao Z, Togano K, Zhang Y, Matsumoto A, Kikuchi A and Kumakura H 2017 High transport $J_c$ in stainless steel/Ag-Sn double sheathed Ba122 tapes *Supercond. Sci. Technol.* **30** 095012

[22] Tamegai T, Pyon S, Miyawaki D, Kobayashi Y, Awaji S, Kito H, Ishida S, Yoshida Y, Takano K, Kajitani H and Koizumi N 2020 Developments of (Ba,Na)Fe$_2$As$_2$ and CaKFe$_4$As$_4$ HIP round wires *Supercond. Sci. Technol.* **33** 104001

[23] Ishida S, Song D, Ogino H, Iyo A, Eisaki H, Nakajima M, Shimoyama J and Eisterer M 2017 Doping-dependent critical current properties in K, Co, and P-doped BaFe$_2$As$_2$ single crystals *Phys. Rev. B* **95** 014517

[24] Jiang J, Bradford G, Hossain S I, Brown M D, Cooper J, Miller E, Huang Y, Miao H, Parrell J A, White M, Hunt A, Sengupta S, Revur R, Shen T, Kametani F, Trociewitz U P, Hellstrom E E and Larbalestier D C 2019 High-Performance Bi-2212 Round Wires Made With Recent Powders *IEEE Trans. Appl. Supercond.* **29** 1–5

[25] Tokuta S and Yamamoto A 2019 Enhanced upper critical field in Co-doped Ba122 superconductors by lattice defect tuning *APL Materials* **7** 111107

[26] Tokuta S, Shimada Y and Yamamoto A 2020 Evolution of intergranular microstructure and critical current properties of polycrystalline Co-doped BaFe$_2$As$_2$ through high-energy milling *Supercond. Sci. Technol.* **33** 094010

[27] Rowell J M 2003 The widely variable resistivity of MgB$_2$ samples *Supercond. Sci. Technol.* **16** R17–27

[28] Senkowicz B J, Mungall R J, Zhu Y, Jiang J, Voyles P M, Hellstrom E E and Larbalestier D C 2008 Nanoscale grains, high irreversibility field and large critical current density as a function of high-energy ball milling time in C-doped magnesium diboride *Supercond. Sci. Technol.* **21** 035009

[29] Yamamoto A, Jiang J, Tarantini C, Craig N, Polyanskii A A, Kametani F, Hunte F, Jaroszynski J, Hellstrom E E, Larbalestier D C, Jin R, Sefat A S, McGuire M A, Sales B C, Christen D K and Mandrus D 2008 Evidence for electromagnetic granularity in the polycrystalline iron-based superconductor LaO$_{0.89}$F$_{0.11}$FeAs *Appl. Phys. Lett.* **92** 252501

[30] Yamamoto A, Polyanskii A A, Jiang J, Kametani F, Tarantini C, Hunte F, Jaroszynski J, Hellstrom E E, Lee P J, Gurevich A, Larbalestier D C, Ren Z A, Yang J, Dong X L, Lu W and Zhao Z X 2008 Evidence for two distinct scales of current flow in polycrystalline Sm and Nd iron oxypnictides *Supercond. Sci. Technol.* **21** 095008

[31] Kim Y-J, Weiss J D, Hellstrom E E, Larbalestier D C and Seidman D N 2014 Evidence for composition variations and impurity segregation at grain boundaries in high current-density polycrystalline K- and Co-doped BaFe$_2$As$_2$ superconductors *Appl. Phys. Lett.* **105** 162604

[32] Li L, Zhang X, Yao C, Liu S, Huang H, Dong C, Cheng Z, Awaji S and Ma Y 2019 Enhancement of the critical current density in Cu/Ag composite sheathed (Ba, K)Fe$_2$As$_2$ tapes by pre-annealing process *Mater. Res. Express* **6** 096003

[33] Lin H, Yao C, Zhang H, Zhang X, Zhang Q, Dong C, Wang D and Ma Y 2015 Large transport J$_c$ in Cu-sheathed Sr$_{0.6}$K$_{0.4}$Fe$_2$As$_2$ superconducting tape conductors *Scientific Reports* **5** 11506

[34] Yao C, Wang D, Huang H, Dong C, Zhang X, Ma Y and Awaji S 2017 Transport critical current density of high-strength Sr1-xKxFe2As2/Ag/Monel composite conductors *Supercond. Sci. Technol.* **30** 075010